\documentclass[aps,twocolumn,a4paper]{revtex4-2}
\usepackage{graphicx}
\usepackage{times}

\usepackage{hyperref}

\begin{document}

\title{Defect-engineering hexagonal boron nitride using low-energy Ar$^+$ irradiation}

\author{Manuel Längle}
\email{manuel.laengle@univie.ac.at}
\affiliation{University of Vienna, Faculty of Physics, Boltzmanngasse 5, 1090 Vienna, Austria}

\author{Barbara Maria Mayer}
\affiliation{University of Vienna, Faculty of Physics, Boltzmanngasse 5, 1090 Vienna, Austria}

\author{Jacob Madsen}
\affiliation{University of Vienna, Faculty of Physics, Boltzmanngasse 5, 1090 Vienna, Austria}

\author{Diana Propst}
\affiliation{University of Vienna, Faculty of Physics, Boltzmanngasse 5, 1090 Vienna, Austria}

\author{Arixin Bo}
\affiliation{University of Vienna, Faculty of Physics, Boltzmanngasse 5, 1090 Vienna, Austria}

\author{Clara Kofler}
\affiliation{University of Vienna, Faculty of Physics, Boltzmanngasse 5, 1090 Vienna, Austria}

\author{Vinzent Hana}
\affiliation{University of Vienna, Faculty of Physics, Boltzmanngasse 5, 1090 Vienna, Austria}

\author{Clemens Mangler}
\affiliation{University of Vienna, Faculty of Physics, Boltzmanngasse 5, 1090 Vienna, Austria}

\author{Toma Susi}
\affiliation{University of Vienna, Faculty of Physics, Boltzmanngasse 5, 1090 Vienna, Austria}

\author{Jani Kotakoski}
\affiliation{University of Vienna, Faculty of Physics, Boltzmanngasse 5, 1090 Vienna, Austria}
\email{jani.kotakoski@univie.ac.at}

% Include the date command, but leave its argument blank.

\date{}

\begin{abstract}
    Monolayer hexagonal boron nitride (hBN) has recently become the focus of intense research as a material to host quantum emitters.
    Although it is well known that such emission is associated with point defects, so far no conclusive correlation between the spectra and specific defects has been demonstrated.
    Here, we prepare atomically clean suspended hBN samples and subject them to low-energy ion irradiation.
    The samples are characterized before and after irradiation via automated scanning transmission electron microscopy imaging to assess the defect concentrations and distributions.
    We find an intrinsic defect concentration of ca. 0.03~nm$^{-2}$ (with ca. 55\% boron and 8\% nitrogen single vacancies, 20\% double vacancies and 16\% more complex vacancy structures).
    To be able to differentiate between these and irradiation-induced defects, we create a significantly higher (but still moderate) concentration of defects with the ions (0.30~nm$^{-2}$), and now find ca. 55\% boron and 12\% nitrogen single vacancies, 14\% double vacancies, and 18\% more complex vacancy structures.
    The results demonstrate that already the simplest irradiation provides selectivity for the defect types, and open the way for future experiments to explore changing the selectivity by modifying the irradiation parameters.
\end{abstract}

\maketitle

\newpage
\section*{Introduction}

Although mostly known as an electrically insulating structural counterpart of graphene, monolayer hexagonal boron nitride (hBN) has recently itself received increasing attention as a solid state host for quantum emitters~\cite{sajid_single-photon_2020}.
What makes it particularly interesting is the good stability of hBN quantum emitters over a wide temperature range~\cite{kianinia_robust_2017}, their bright emission into the zero-phonon line~\cite{tran_quantum_2016}, wide spectral range~\cite{tran_robust_2016,bourrellier_bright_2016, gottscholl_initialization_2020}, and the possibility for lifetime-limited emission at room temperature~\cite{hoese_mechanical_2020,dietrich_solid-state_2020}.
This makes hBN attractive for a number of different advanced applications~\cite{de_leon_materials_2021}.
However, although it is well known that quantum emitters are associated with point defects in the material~\cite{tran_robust_2016,bourrellier_bright_2016,su_tuning_2022,wong_characterization_2015,abdi_color_2018}, no direct correlation between the defect structures and quantum emission properties has been established.
Therefore, being able to selectively create specific types of point defects into hBN would be desirable.

It has been already shown that different kinds of irradiation can be used to create quantum emitters in hBN, including lasers~\cite{gan_large-scale_2022}, ions~\cite{choi_engineering_2016}, neutrons~\cite{zhang_discrete_2019} and electrons~\cite{exarhos_optical_2017,su_tuning_2022}.
So far only in the case of electrons~\cite{bui_creation_2023} the direct correlation between irradiation and the exact atomic structure has been established.
In the case of ion irradiation, analytical potential molecular dynamics simulations~\cite{lehtinen_production_2011,ghaderzadeh_atomistic_2021} have suggested that low-energy noble gas ions should lead to formation of point defects with a high likelihood (40--80\% per ion) and similar probabilities for single boron and nitrogen vacancies.
Especially according to simulations~\cite{ghaderzadeh_atomistic_2021} with the improved potential benchmarked against density functional theory, the best selectivity should be achievable at the lowest energies ($<$200~eV), where nitrogen single vacancies become clearly more favorable for projectiles heavier than Ne.
However, this prediction has not been confirmed experimentally.
Moreover, electron irradiation experiments~\cite{bui_creation_2023} have shown that it is more likely to displace boron than nitrogen, also at intermediate energies (60--100~kV) where inelastic interactions are significant.
Whether these need to be taken into account also for low-energy single-charged ion irradiation of hBN remains unclear.

Here, we prepare atomically-clean suspended samples from commercial monolayer hBN grown via chemical vapor deposition and irradiate them with low-energy Ar$^+$ ions.
The intrinsic defect density and distribution are measured using automated scanning transmission electron microscopy (STEM) annular dark field (ADF) imaging at 60~kV in ultra-high vacuum.
We find a fairly high intrinsic defect density, which is approximately constant within each sample.
There is a high prevalence of single boron vacancies $V_\mathrm{B}$ (55\%) followed by single nitrogen vacancies $V_\mathrm{N}$ (8\%).
The defect concentration after irradiation is selected to be ca. ten times higher than the intrinsic one on the one hand to be high enough to clearly see the difference between intrinsic and irradiation-induced defects, but on the other hand low enough so that the vast majority of impinging ions encounter pristine non-defective hBN.
Also the post-irradiation defect distribution is dominated by boron single vacancies (55\%), followed by nitrogen single vacancies (12\%), double vacancies (14\%) and more complex vacancy defects (18\%).
These results show that some defect-selectivity indeed can be achieved with low-energy noble gas irradiation.
However, experimentally the most common defect is the boron single vacancy in contrast to the prediction from molecular dynamics.
Future work should explore whether modifying irradiation parameters (ion species, kinetic energy, charge state, irradiation angle) changes the distribution either from boron to nitrogen single vacancies or more complicated defect structures.

\section*{Results and discussion}

\begin{figure}[t]	
	\centering
	\includegraphics{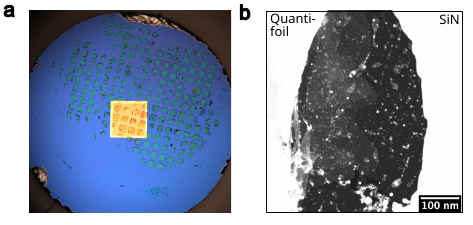}
    \caption{{\bf Overview of the sample.}
		(a) Light microscopy image of the sample after transfer.
        The purple/blue area corresponds to the Si grid, the light blue squares to Quantifoil.
        The yellow square is the perforated SiN window where the suspended sample can be found when holes in SiN and the Quantifoil overlap.
        (b) STEM-MAADF image of a suspended sample area.
        The dark contrast is clean suspended hBN and the gray areas correspond to fewlayer hBN (continuous areas on the left-hand side and to contamination smaller areas throughout the sample).
        The white area on the left corresponds to Quantifoil and the white area on the right to SiN.
	}
	\label{fig:methods}
\end{figure}

Commercial hBN grown via chemical vapor deposition was transferred onto a Au Quantifoil TEM grid as described~\cite{ahmadpour_monazam_substitutional_2019,bui_creation_2023}.
The Quantifoil membrane with hBN was subsequently further transferred onto a Si grid with a perforated SiN membrane by mechanically delaminating it from the Au grid to create a hybrid Quantifoil SiN grid (see Methods).
A light microscopy image of a transferred sample is shown in Fig.~\ref{fig:methods}a.
This hybrid grid allows a polymer-free transfer of hBN onto the final grid, which is mechanically robust and allows sample cleaning with a laser in vacuum~\cite{tripathi_cleaning_2017}.
After preparation, the samples were inserted into an extensive vacuum system~\cite{mangler_materials_2022} that connects all instruments used in this study through high-vacuum transfer lines (base pressure typically below 10$^{-8}$ mbar).
Upon insertion, the samples are baked for ca. 10~h at 150$^{\circ}$C in vacuum to remove water and the most severe contamination.
The sample is surveyed via scanning transmission electron microscopy (STEM) medium angle annular dark field (MAADF) imaging using the Nion UltraSTEM 100 in Vienna at 60~kV under ultra-high vacuum conditions to find a location with good coverage and reasonable surface cleanliness.
After this, a $2$~ms laser pulse is used to remove most of the remaining contamination (see Methods).
%As a control experiment, one sample instead of the laser pulse cleaned by vacuum heating at 500$^\circ$~C for 1~h.
An example large-scale STEM-MAADF image of a cleaned sample is shown in Fig.~\ref{fig:methods}b.

After cleaning, we carry out automated atomic-resolution STEM-MAADF imaging~\cite{mittelberger_automated_2017} to estimate the intrinsic defect concentration and distribution.
This method allows collecting a large amount of data, and in combination with automated machine-learning based analysis enables minimizing the electron dose that the sample is exposed to.
A total of five samples were prepared and pre-characterized in this manner, and one sample was selected for ion irradiation.
In total, 16500~nm$^2$ of the selected sample was imaged, out of which ca. 68\% was contamination-free and imaged with a resolution that allows atomic structural analysis.
The images are analyzed automatically with a convolutional neural network using a pipeline similar to those described in Refs.~\cite{ziatdinov_deep_2017,trentino_atomic-level_2021-1} (see Methods for details).
In short, the neural network first separates image areas to either contamination or lattice.
Next, each lattice site is assigned an intensity value at different values of Gaussian blurring.
These values are provided to a classifier that determines whether the site contains an atom or a vacancy.
The process is illustrated in a previous publication~\cite{trentino_atomic-level_2021}.
%Fig.~\ref{fig:analysis}.
In total, 335 defects were identified in the data set recorded with the sample selected for irradiation, corresponding to a defect concentration of ca. 0.03~nm$^{-2}$ with 55\% boron and 8\% nitrogen single vacancies.
%
%\begin{figure*}[b!]	
%	\centering
%	\includegraphics[width=0.7\textwidth]{figures_tmp/neural_network_figure.pdf}
%	\caption{{\bf Automated analysis pipeline.}
%    The analysis pipeline operates on a stack of images.
%    Each image is analyzed with a convolutional neural network to find the (possible) contamination and each of the hBN lattice sites.
%    For each lattice site, local intensity features are extracted, which are used to classify whether the site contains an atom or a vacancy.
%    In addition, the geometric relationship between the lattice sites is used for finding neighboring lattice sites and separating the boron and nitrogen sublattices.
%    Finally, the defect size and type is determined by counting connected vacancies.
%	}
%	\label{fig:analysis}
%\end{figure*}

It is worth pointing out, that electron irradiation during imaging is known to introduce defects into hBN~\cite{meyer_selective_2009,jin_fabrication_2009,kotakoski_electron_2010} even under ultra-high vacuum~\cite{bui_creation_2023}.
While recording the images we should have created only ca. 29.7$ \pm$2.8 defects (of which 59\% should be boron and 41\% nitrogen vacancies), taking the typical beam current close to 20pA of our instrument~\cite{speckmann_combined_2023} and the imaging parameters~\cite{bui_creation_2023} (see Methods). Therefore, ca. 90\% of the defects must be intrinsic to the sample and arise from growth or sample fabrication, leading to an intrinsic defect concentration of ca. 0.027~nm$^{-2}$.

\begin{figure}[]	
	\centering
	\includegraphics{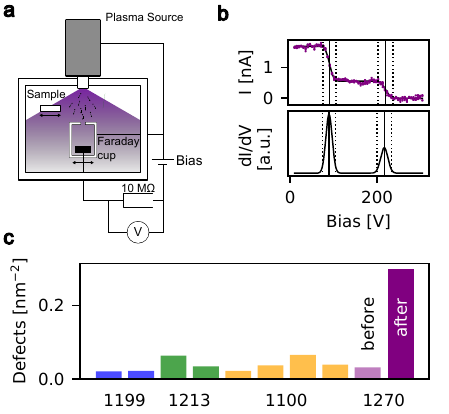}
	\caption{{\bf Plasma setup, beam profile measurement and defect concentration.}
        (b) Schematic presentation of the plasma irradiation setup and the Faraday up.
        (d) Measured ion current $I$ and the calculated beam profile ($\mathrm{d}I/\mathrm{d}V$) as a function of the bias voltage ($V$).
        The purple points are measured data, and the black line correspond to a fit (top) and its derivative (bottom).
        (c) Defect concentration for pre-characterized samples (1199 in blue, 1213 in green, 1100 in yellow and 1270 in purple) compared to the irradiated one (1270 in purple).
        All samples were cleaned by a laser pulse before imaging.
	}
	\label{fig:methods_plasma}
\end{figure}

Next, the selected sample was transported in vacuum to a chamber containing a plasma generator, where low-energy Ar$^+$ ion irradiation is used to create defects (Fig.~\ref{fig:methods}a) following Ref.~\cite{trentino_atomic-level_2021-1}.
With the parameters used for the plasma source (see Methods), the beam profile (Fig.~\ref{fig:methods_plasma}b), measured with a biased Faraday cup, reveals two peaks: one at a low energy of 228 $\pm$ 11~eV close to the manufacturer specification, and another one at 89 $\pm$ 10~eV.
Peak intensities measured over several experiments show that the lower-energy peak contains ca. 2/3 of all ions.
The estimated total irradiation dose, based on current measurement with the Faraday cup and taking pressure variations and the geometry of the setup, is ca. 0.10 $\pm$ 0.04~ions/nm$^2$.

After ion irradiation, the sample was further automatically imaged to estimate the post-irradiation defect density and distribution at another sample position.
A comparison between the defect concentration in non-irradiated and irradiated samples is shown in Fig.~\ref{fig:methods_plasma}c, 
As is obvious from the data, the non-irradiated samples have a non-negligible defect concentration, which varies from sample to sample, but remains fairly similar within each of the characterized samples, shown in Fig.~\ref{fig:methods_plasma}c.
These values are significantly lower than those reported in Ref.~\cite{wang_photoluminescence_2018}, which may be explained by the neglect of electron-beam induced damage in the earlier study and the additional damage-inducing effect of a non-ultra-high vacuum atmosphere in their microscope column~\cite{javed_hBN_2024}.

\begin{figure}[b]	
	\centering
    \includegraphics{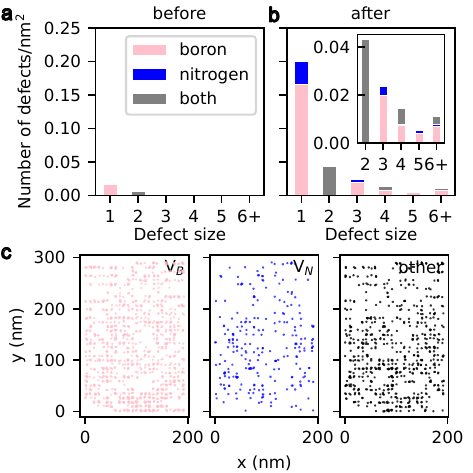}
	\caption{{\bf Defect concentration and distribution.}
        (a) Intrinsic defect concentration as a function of the defect size (number of missing atoms).
        11,242~nm$^2$ of valid area were analysed before and 7,246~nm$^2$ of valid area were analysed after the irradiation.
        (b) Post-irradation defect concentration as a function of the defect size.
		The inset shows a magnified view of the defect concentration for defects with at least two missing atoms.
        (c) Spatial distribution of boron and nitrogen single vacancies ($V_\mathrm{B}$ and $V_\mathrm{N}$, respectively) and all other defects.
	}
	\label{fig:defects_figure}
\end{figure}

\begin{figure}[t]	
	\centering
	\includegraphics{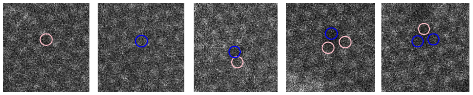}
	\caption{{\bf Example STEM-MAADF images of the most typical defects.}
		Pink circles mark the positions of missing B and blue circles missing N atoms.
        From left to right: $V_\mathrm{B}$, $V_\mathrm{N}$, double vacancy, triple vacancy with one missing N and two missing N, and triple vacancy with one missing B and two missing N.
	    All images have a field of view of 2$\times$2~nm$^2$.
    }
	\label{fig:defect_types}
\end{figure}

Defect distributions for the irradiated sample before and after irradiation are shown in Fig.~\ref{fig:defects_figure}a,b.
After irradiation, a total area of ca. 14,500~nm$^2$ was imaged with ca. 50\% that could be used for the analysis (clean and sufficient resolution).
In total, 2169 defects were identified, leading to a defect concentration of ca. 0.30~nm$^{-2}$, i.e., close to ten times higher than before the irradiation.
The amount of irradiation-induced defects is thus ca. 0.27~nm$^{-2}$.
We point out that the defect density is also higher than the estimated total irradiation dose (by a factor of nearly three).
This most likely reveals the inherent uncertainty in estimating accurately the current of the plasma source by measurements carried out separately from the irradiation experiment.
The defect density is low enough so that the vast majority of the ions have impinged on pristine defect-free hBN.
After the irradiation, ca. 55.6\% of defects are boron and 11.5\% nitrogen single vacancies.
Further 14.4\% are double vacancies and the rest (18.4\%) more complex vacancy defects.
The spatial distributions for $V_\mathrm{B}$ and $V_\mathrm{N}$ as well as all other defects are shown in Fig.~\ref{fig:defects_figure}c.
The grid-like pattern of the defect locations arises from the images recorded in discrete non-overlapping parts of the sample and the larger defect-free areas have been obscured by surface contamination.
Overall, the spatial distribution shows that the defects have been created uniformly on the sample.
Example STEM-MAADF images of the most typical defects are shown in Fig.~\ref{fig:defect_types}.

%A study conducted by \cite{wang_photoluminescence_2018} aberration-corrected JEOL ARM-200F 
%%exfoliated monolayer h-BN with high defect densities
%STEM studiy - defects, STEM-ADF, 60~kV, (claim it is lower than the knock on threshold of 78 kV)
%V$_B$ 0.11 nm$^{-2}$
%V$_N$ 0.0047 
%V$_{BN}$ 0.0031
%V$_{B3N}$ 0.022

%V$_B$ = 78.7\%
%V$_N$ = 3.4 \%
%V$_{BN}$ = 2.2 \%
%V$_{B3N}$ = 15.7 \%

%investigated 12800 nm$^2$

%used DFT calculation to see which defect is responsible for emission, make no actual statement, do not tell about the effect of the electron beam on the defects
%V$_B$ and $V_N$ are identified as possible candidates

Interestingly, the results are in a clear contradiction with theoretical studies based on analytical-potential molecular dynamics simulations, where it was found depending on the potential used that either low-energy Ar irradiation should lead to about 17\% single vacancies of B and  12--17\% single vacancies of N~\cite{lehtinen_production_2011} per impinging ion or that the probability would be between 35--55~\% for B and 40--65~\% for N~\cite{ghaderzadeh_atomistic_2021}  for energies similar to the ones we determined experimentally.
On the experimental side, since we do not have an accurate measurement of the ion current during the irradiation, we can not precisely know what the probability for creating each type of single vacancy is. 

However, since we observe about three times as many defects as we estimate ions to have irradiated the sample, it seems a safe assumption that the probability must be closer to one than the simulations suggest.
Additionally, carbon atoms from the ubiquitous contamination are known to migrate on the surface of 2D materials and fill vacancies~\cite{inani_silicon_2019}, especially at elevated temperatures~\cite{postl_indirect_2022}.
Due to the similar atomic number (and therefore contrast~\cite{krivanek_atom-by-atom_2010}) of C as that of B and N, it is possible that our characterization has missed some vacancies that have been subsequently filled with carbon.
However, here the same argument applies as above: since the defect concentration is higher than expected based on the estimated ion current, it seems unlikely that such a large number of N vacancies would have been filled with carbon atoms that the experimental results would agree with the simulation prediction.
It is typically assumed~\cite{lehtinen_effects_2010} that the singly-charged ion impact can be modelled without taking charges into account.
This is because upon an impact the ion receives an electron from the sample neutralizing.
While for conductive samples the missing charge distributes quickly through the material, this is likely not true for an insulating 2D material such as hBN where the charge may remain localized in the vicinity of the impact point.
This leads to a decrease in the displacement threshold energy.
Therefore, the most likely explanation seems to be that analytical molecular dynamics is not a sufficiently accurate method to describe the ion irradiation process at the lowest energies.

In accordance with the higher likelihood for obtaining boron vacancies, also the displacement threshold as calculated with density functional theory-based molecular dynamics for a neutral hBN lattice is lower for boron than for nitrogen (19.36~eV vs. 23.06~eV, respectively)~\cite{kotakoski_electron_2010}.
However, since energy transfer from Ar to N is more efficient than to B due to the higher mass of N, this fact by itself is not sufficient to explain our experimental results, and the complete description may require including effects due to charging or inelastic scattering as alluded to above.
Nevertheless, assuming that the lower energy ions of our beam (ca. 90~eV) would nearly exclusively produce boron vacancies, whereas the higher energy ions (ca. 230~eV) would produce both vacancies with a similar probability that would lead to a defect distribution similar to the observed one.
Overall, this discrepancy highlights that more experimental and computational work is required to obtain a comprehensive understanding of the capabilities of this method for defect-specific engineering of hBN.

\section*{Conclusions}

We demonstrated that low-energy Ar$^+$ irradiation is a suitable method for defect-engineering hBN with a high selectivity for boron single vacancies.
The intrinsic defect concentration in the as-prepared and cleaned samples was found to be below 0.27~nm$^{-2}$ (with 55\% boron and 8\% nitrogen vacancies).
Additional ca. 0.003 defects per nm$^2$ was estimated to have been created during the automated imaging of the sample pre-irradiation.
The irradiation dose was chosen high enough to separate the irradiation-induced defects from the intrinsic ones, but sufficiently low to avoid ions impinging on already defective sites.
Most (ca. 55.6\%) of the irradiation-induced defects were boron single vacancies, followed by nitrogen single vacancies (11.5\%), double vacancies (14.4\%) and more complicated vacancy structures.
These results are in contrast to earlier computational work predicting either similar probabilities for B and N single vacancies or a prevalence of N vacancies as a result of Ar$^+$ irradiation at similar energies.
This shows clear promise for selectively creating specific defects with low-energy ions, but also that more research is needed both experimentally and computationally to fully exploit the possibilities for defect-engineering hBN with low-energy ions.

\bibliography{scibib}
\bibliographystyle{naturemag_doi_jk}
%\bibliography{references}

\section*{Methods}

\subsection*{Sample preparation and cleaning}

Suspended monolayer hBN samples were fabricated from commercial hBN grown via chemical vapor deposition (Graphene Supermarket).
First, hBN was transferred from the Cu growth substrate onto a Au Quantifoil TEM grid (a periodicity of 2~$\mathrm{\mu m}$ and a hole size of 2~$\mathrm{\mu m}$) following the procedure described in Refs.~\cite{ahmadpour_monazam_substitutional_2019,bui_creation_2023} (Cu was etched in a bath of $1.5$~g of iron chloride mixed with $240$~g deionized water for 47~h, after which the sample was washed in three cycles of water and isopropyl alcohol, for ca. 1~min each).
Then, hBN on Quantifoil was placed onto a Si grid with a perforated SiN window (hole size of 1~$\mu$m) where it was being adhered by putting a small drop of isopropyl alcohol on it and baking at 150$^\circ$C for 15~min and after that it was mechanically delaminated from the Au grid with a tweezer.
The suspended sample used for the experiments can be found where the holes of the Quantifoil and the SiN overlap.

Upon being inserted into the vacuum system used for all experiment, the samples are baked for ca. 10~h at 150$^\circ$C in vacuum.
After precharacterization to find a suitable sample area, it is cleaned by applying a laser pulse in the column of the scanning transmission electron microscope used for imaging.
Due to an incidence angle of 25$^\circ$, the laser spot has the shape of an ellipse with a main axis of 28 $\pm$ 3~$\mu$m and a minor axis of 35 $\pm$ 3~$\mu$m.
The loss in the optical system is about 66\%, which means that the 2~ms and 10~mW pulse of a continuous wave laser used for cleaning corresponds to ca. 6.8~$\mathrm{{\mu}}$J with an energy density of  2150$\pm $70~Jm$^{-2}$ or 13400 $\pm$ 400~eVnm$^{-2}$.

\subsection*{Ion irradiation}

The samples were irradiated with an low-energy Ar$^+$ beam using a SPECS ECR-HO microwave plasma source operated in hybrid mode with an anode voltage of 0~V and extractor voltage of -60~V.
The pressure of the chamber was kept approximately constant at ca. $3.5\times 10^{-6}$~mbar, though variations from this could not be entirely avoided, and the irradiation time was ca. $150$~s.
The irradiation is carried out at room temperature, similar to all other the experiments except for sample cleaning.
The current was measured with a Faraday cup placed in the same place as the sample with a slightly larger distance, which due to the beam divergence leads to a difference in the measured current and the current at the sample of approximately a factor of $1.4$ (current at the sample being higher).
To get an accurate estimate of the ion energies, the current was measured as a function of the bias voltage, where the ions appear as drops in the measured current that deviates from the quadratically decaying background current. % determined by the resistivity of the measurement system.
This reveals that the ion beam consists of two distinct Gaussian distributions, one centered at 89 $\pm$ 10~eV and the other one at 228 $\pm$ 11~eV, the lower-energy peak contributing ca. 2/3 of the ions.
The total current was calculated by summing up the ions contributing to each of these distributions.
From the ion current and the irradiation time, we estimate a total dose at the sample to be 0.10 $\pm$ 0.04 ions per nm$^2$.

\subsection*{Microscopy}

All electron microscopy images were recorded with the Nion UltraSTEM 100 microscope in Vienna at 60~kV in ultrahigh vacuum (10$^{\mathrm{-10}}$~mbar) using the medium angle annular dark field (MAADF) detector with annular ranges of 60--200~mrad.
The convergence semi-angle was 30~mrad, and the beam current in this mode is typically about 20~pA.

For obtaining defect concentrations and distributions, we used automatic image acquisition~\cite{mittelberger_automated_2017}.
With this method, the algorithm records images in a predefined area of the sample along a serpentine path.
Here, we defined the grid of recorded images to avoid overlap between the images by leaving an empty area of a similar size as the images between each pair of subsequent images.
During the automatic image acquisition minor adjustments of the electron energy and the correction of astigmatism and coma were done manually.
The images were recorded with a (calibrated~\cite{madsen_fourier-scale-calibration_2022}) field of view of 4.2$\times$4.2~$\mathrm{nm}^2$, a pixel dwell time of 16~$\mathrm{\mu s}$ and 512$\times$ 512 pixels.

Taking the imaging parameters, the flyback time at the end of each scan line (120~$\mu$s) and half of the time to move the stage from one position to the next (0.5$\times $2~s), the electron dose per frame is (3.6 $\pm$ 0.14) $\times$ 10$^5$~electrons \AA$^{{-2}}$ using a current of $19.4 \pm 0.6$~pA~\cite{speckmann_combined_2023}.
Since there are ca. 36.8~atoms nm$^{-2}$, with the correct displacement cross section~\cite{bui_creation_2023}, we estimate that ca. 0.046 $\pm$ 0.005 defects are created per image (or 0.0026 $\pm$ 0.0003 per nm$^2$, of which 0.0016 $\pm$ 0.0003 are boron and 0.0011 $\pm$ 0.0005 per nm$^2$ nitrogen).
Therefore, ca. 29.7 $\pm$ 2.8 (out of 335) defects pre-irradiation and 19.1 $\pm$ 1.9 (out of 2169) defects post-irradiation can be estimated to have been created during imaging.

\subsection*{Automated image analysis}

The convolutional neural network implemented in Pytorch~\cite{NEURIPS2019_9015} has a UNET structure with rotational equivariance~\cite{ronneberger_unet_2015,weiler_equivariant_2019}, and its architecture is identical to that described in Ref.~\cite{trentino_atomic-level_2021}.
The neural network has two output branches producing a segmentation map of contaminated areas and a density map of lattice sites.
The density map has a high value at every lattice site, whether an atom is present or there is a vacancy.
This was found to be more robust than training the network to detect the defects directly.
The neural network is trained from simulated data (as in Ref.~\cite{trentino_atomic-level_2021}).
For the simulated data, randomized atomic models of hBN are generated, and the models are used as input for the multislice algorithm to simulate annular dark field images~\cite{madsen_abtem_2021}.
Finally, random distortions and noise are applied to the simulated images.

The neural network is strictly trained for images with a pixel size of 0.1~\AA, and therefore images with different pixel sizes are resized.
The scaling factor is determined by finding diffraction spots in the Fourier-transformed images.

Given the density map of lattice sites, a discrete set of points representing the detected lattice is extracted.
Whether a lattice site represents a vacancy is determined by looking at the local intensity at varying levels of Gaussian blurring.
Given these intensity features, a small random forest classifier is trained from twenty hand-labeled experimental images.
The trained classifier is then applied to the whole set of images to label every lattice site as a vacancy or a non-vacancy.

In order to arrive at the final classified defects, a geometric analysis of the lattice sites is also performed.
First, a geometric graph is created by connecting neighboring lattice sites.
A graph coloring algorithm separates the two sublattices; each sublattice may then be assigned to boron or nitrogen based on local intensity. 
A local lattice segment comprised of each point and its three neighbors in the graph are matched to a template representing the expected local geometry using the quaternion characteristic polynomial method~\cite{theobald_qcp_2005}.
The root-mean-squared distance between the expected and measured local lattice is used to measure how well each local segment of the lattice matches an ideal hexagonal lattice.
Lattice segments that deviate too much are thrown away from the analysis, as they indicate severe scan distortions, noise, or highly distorted lattice areas. 
Given each labeled lattice site and the geometric graph, the defects are classified based on how many boron and nitrogen atoms they contain by finding the connected vacancy sites.

%\bibliography{scibib}
%\bibliographystyle{Science}
\bibliographystyle{naturemag_doi_jk}
%\bibliography{references}

\clearpage

\setcounter{figure}{0}
\makeatletter 
\renewcommand{\thefigure}{S\@arabic\c@figure}
\makeatother

%\section*{Supplement}

%Data selection:
%only take statistically relevant datasets (>50 images)
%
%in 5 samples we saw relatively little spread in defect concentration within one sample and always defect densities which were at least half of what we saw after the irrad
%
%1199 - 108 + 98 images - 815 + 918 nm$^2$ - around 3-5 defects/nm$^2$
%1213 - 58 (501 nm$^2$) - around 0.13-0.15 , 66 (1080 nm$^2$) - around  0.13-0.15, 196 (2297 nm$^2$) - around 0.11-0.13 , 126 (2144 nm$^2$) - around 0.12-0.14 
%1100 - 76 (820 nm$^2$) - below 0.1,  56(674 nm$^2$) - below 0.1,  156 (1745 nm$^2$) - below 0.08 , 222 (2435 nm$^2$) - below 0.08
%1270 - 103 (1828 nm$^2$) - below 0.13, 753(9072 nm$^2$) - below 0.07
%1270 after irrad - 603(5967 nm$^2$) - about 0.35 defects/nm$^2$
%1890 heated - 61 (657 nm$^2$) - below 0.07, 94 (1054 nm$^2$) - below 0.04

\end{document}